\documentclass[conference]{IEEEtran}
\IEEEoverridecommandlockouts
\usepackage{cite}
\usepackage{amsmath,amssymb,amsfonts}
\usepackage{algorithmic}
\usepackage{graphicx}
\usepackage{textcomp}
\usepackage{xcolor}
\def\BibTeX{{\rm B\kern-.05em{\sc i\kern-.025em b}\kern-.08em
    T\kern-.1667em\lower.7ex\hbox{E}\kern-.125emX}}
\begin{document}

\title{Simplified EPFL GaN HEMT Model\\
\thanks{This project is funded by the Swiss National Science Foundation - project 200021 213116.}
}

\author{Farzan Jazaeri, Majid Shalchian, Ashkhen Yesayan, Amin Rassekh, Anurag Mangla, \\ Bertrand Parvais, and Jean-Michel Sallese
\thanks{Farzan Jazaeri, Ashkhen Yesayan, and Jean-Michel Sallese are with the Electron Device Modeling and Technology Laboratory (EDLAB) of the École Polytechnique Fédérale de Lausanne (EPFL), Switzerland (e-mail:farzan.jazaeri@epfl.ch).Majid Shalchian is with the department of Electrical Engineering, Amirkabir University of Technology. Amin Rassekh is with InCize, Louvain-la-Neuve, Belgium. Anurag Mangla is an alumnus of EPFL. 
Bertrand Parvais is affiliated with IMEC in Leuven and holds a position as a Guest Professor at Vrije Universiteit Brussels, Belgium.}\vspace{-0.5cm}}

\maketitle
\begin{abstract}
This paper introduces a simplified and design-oriented version of the EPFL HEMT model\cite{HEMT}, focusing on the normalized transconductance-to-current characteristic ($ Gm/I_D $). Relying on these figures, insights into GaN HEMT modeling in relation to technology offers a comprehensive understanding of the device behavior. Validation is achieved through measured transfer characteristics of GaN HEMTs fabricated at IMEC on a broad range of biases. 
This simplified approach  should enable a simple and  effective circuit design methodology with AlGaN/GaN HEMT heterostructures.
\end{abstract}
\IEEEpeerreviewmaketitle
\section{Introduction}

High Electron Mobility Transistors (HEMTs) have garnered interest due to their exceptional electron transport capabilities boosted by a quantum well that enables high-speed and high-power applications in microwave and millimeter-wave domains.

Despite these advantages, existing compact models and traditional design methods with HEMTs rely on outdated interpretations of their behavior which limits hand calculations in a preliminary design phase. To this purpose, we introduce the concept of inversion coefficient introduced initially in MOSFETs \cite{EKV1}, combined with the important $ G_m/I_D $ figure of merit. This approach leads to a comprehensive set of analytical expressions based on the charge-voltage relationships outlined in \cite{HEMT, tran, 9137643} and valid in all the regions of operation. Likewise for the MOSFET, we aim at giving a simple and effective design-oriented modelling framework.

The simplified model relies only on six model parameters:
the slope factor $ n_q $, the threshold voltage $ V_{T0} $, the specific current $ I_{sp} $, the velocity saturation coefficient $ \lambda_{c} $, the mobility reduction coefficient $ \theta $, and the charge threshold voltage, $ V_T $. This leads to a general modeling and designing methods with HEMTs since the model inherits from the regular silicon MOSFET features, but with new physical quantities, a kind of generalized \textit{MOSFET}. Then, the design strategy adopted in CMOS circuits can be seamlessly applied to the design of circuits based on HEMTs.

\section{Design-oriented Modeling in HEMTs}
To extract analog design parameters that are used to predict the electrical characteristics of HEMTs, we propose to simplify the EPFL HEMT model making use of the $ G_m/I_D $ methodology. This model relies on semiconductor physics, making it a valuable tool for optimizing circuit designs with GaN with few model parameters.

\subsection{Charge-Voltage Dependence}
The typical structure of a HEMT is shown in Fig. 1. It consists in a large bandgap AlGaN semiconductor playing the role of the confinement barrier combined with a smaller bandgap GaN semiconductor.
For simplicity we propose to reuse the explicit and continuous charge-based relationship developed in \cite{HEMT,jazaeri2017free} that we recal hereafter
\begin{equation} \label{1}
\!\!U_T\ln\left[\exp\left(\frac{n_{ch}}{DoS_{2D} U_T}\right)\!-\!1\right]\!+\!\frac{qn_{ch}}{C_{b} n_{q}}  +u_{1}\sqrt[3]{n_{ch}^2}\!=\!\psi_{p}-V,
\end{equation}
where $V= U_T\ln\left(N_A/N_V\right)+E_{g}/q+V_{ch} $, $ u_{1}=\Gamma/q$, $  C_{b}=-\varepsilon_{1}/x_{1} $ where $ x_{1} $ is the AlGaN layer thickness, $U_T $ is the thermal voltage ($ k_BT/q $), $ E_g $ is the band gap of GaN and the temperature-dependency of $ E_g $ is described in \cite{EG}:
\begin{equation}\label{2a}
E_g(T)=E_g(0)-5.08\times 10^{-4} T^2/(996-T),
\end{equation}
where $ E_g (0) = 3.28eV $ for Zinc Blende crystal structure, and other symbols are defined in Table \ref{Table2}.
As initially introduced in \cite{linearization} for MOSFETs, then adopted in HEMTs \cite{HEMT}, the pinch-off surface potential $ \psi_p $ is defined as the surface potential $ \psi_s $ when $ n_{ch}=0 $  
\begin{equation}\label{01}
Q_{ch}=-qn_{ch}=C_b(V_{GB}+V_A+\psi_s+\gamma \sqrt{\psi_s} ),
\end{equation}
leading to
\begin{equation}\label{2}
\!\!\psi_{p}\!=\!V_{GB}\!-\!V_{A}\!-\!\gamma^{2}\left(\sqrt{\frac{V_{GB}\!-\!V_{A}}{\gamma^{2}}\!+\!\frac{1}{4}}\!-\!\frac{1}{2}\right)\,
\end{equation}
where the body factor $\gamma  $ and the voltage offset $ V_A $ are given by
\begin{equation}\label{3}
\gamma=-\frac{x_{1}}{\varepsilon_{1}}\sqrt{2\varepsilon_{2}N_{A}q},
\end{equation}
\begin{equation}\label{4}
V_{A}=-\frac{\Delta E_{C}}{q}+\phi_{B,eff}-\frac{qN_{D}}{2\varepsilon_{1}}x_{1}^{2}.
\end{equation}
\begin{figure}[t]
	\centering
	\includegraphics[width=0.8\columnwidth]{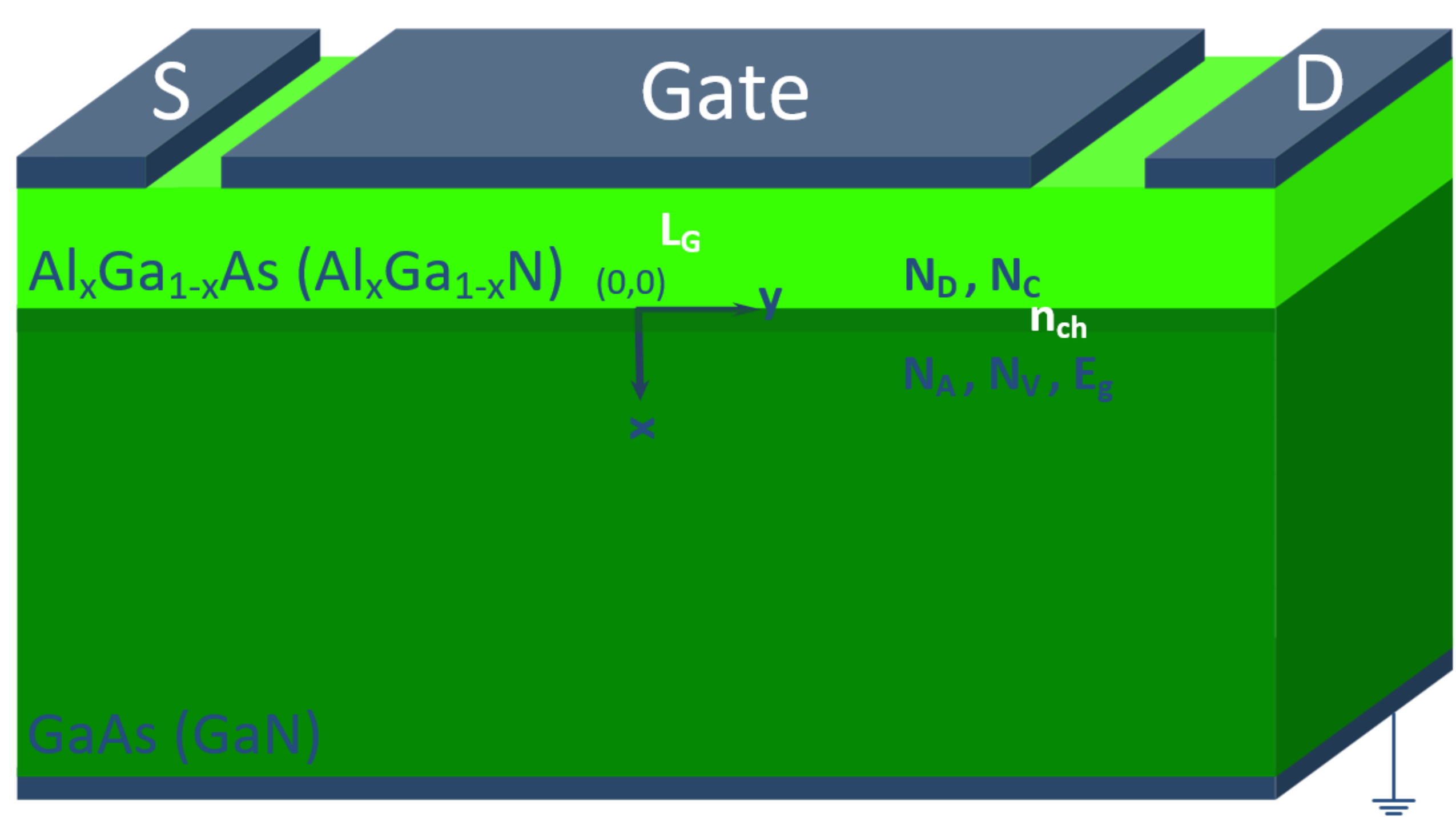}
	\caption{The 3D schematic view of HEMT. The AlGaAs (AlGaN) and GaAs (GaN) regions are respectively \textit{n}-doped with $ N_D $ and \textit{p}-doped with $ N_A $.}
	\label{Fig1}
	\vspace{-0.4cm}
\end{figure}
The effective Schottky barrier height $\phi_{B,eff}$  accounts for polarization effects. Its value is given by $ \phi_{B_{eff}}=\phi_{B}+x_1\sigma_{pol}/\varepsilon_1 $ where $  \sigma_{pol} =\sigma_{pol}(GaN)- \sigma_{pol}(AlGaN) $.  The terms $ \sigma_{pol}(GaN) $ and $ \sigma_{pol}(AlGaN) $ are the polarization-induced charge surface densities localized at the GaN-AlGaN interface. The slope factor $n_q$ is defined as the derivative of $n_{ch}$ versus $ \psi_s $, obtained from relation \ref{01}, evaluated at $\psi_p$ (\cite{HEMT,linearization}),  leading to
$ n_{q}=1+\gamma/2\sqrt{\psi_{p}} $. This slope factor is part of the linearization scheme that links the mobile charge density to the surface potential (\cite{HEMT}, \cite{linearization}):
\begin{equation}\label{linearization}
Q_{ch}=-n_{q} C_{b} (\psi_p-\psi_s),
\end{equation}
where $C_{b}$ holds for an equivalent gate stack capacitance. We can also use the first order term of the Taylor series of (\ref{2}) with respect to $V_{GB}$ to get an approximate value for $\psi_p$, $ \psi_p\approx (V_{GB}-V_A)/n_q $ (\ref{2}). Replacing this value in (\ref{1}) leads to the following  normalized charge-based expression  
\begin{equation} \label{x}
\ln\left[\exp\left(\alpha q_{ch}\right)-1\right]+2q_{ch}  + \beta\sqrt[3]{q_{ch}^2}=\frac{V_{GB}-VT_0}{n_qU_T},
\end{equation}
where $ \alpha =-Q_{sp}/qDoS_{2D}U_T $, $ \beta=(U_1/U_T)\sqrt[3]{(-Q_{sp}/q)^2} $, $ Q_{sp} $ is the specific charge density defined as  $ Q_{sp} = -2n_qC_bU_T $ and $ V_{T0} $  is the threshold voltage defined as  
\begin{equation}
V_{T0}=V_A+n_qU_T\ln\left(N_A/N_V\right)+n_qE_{g}/q+n_qV_{ch}. 
\end{equation}
\subsection{Drain to Source Current Derivation in HEMTs}
The drain current expression of the EPFL HEMT model is given by $ I_{ds}=I_{sp}\times i $ where $ i $ is the normalized current and $  I_{sp} $ is the so-called specific current defined as $ I_{sp} = 2n_q\mu_0 C_bU^2_T W/L_G $. Here, $ i $ is obtained by 
\begin{equation}\label{current}
	i =(q^2_s+q_s-q^2_d+q_d)=i_f-i_r,
\end{equation}
where  $ q_s =-q n_{ch,s}/Q_{sp} $ and $ q_d=-q n_{ch,d}/Q_{sp} $ are the normalized charge densities at source and drain, and $ i_f $ and $ i_r $ are defined as the forward and reverse normalized currents. Each of them is related to the normalized charge densities $ q_s $ and $ q_d $ by $ i_{f,r} =q_{f,r}^2+q_{f,r} $.
These expressions make the HEMT model close the MOSFET \cite{HEMT, linearization} and motivate their use for circuit design. %Therefore $ q_{f,r} $ can be given by 
%$ q_{f,r} =\sqrt{i_{f,r}+\frac{1}{4}}-\frac{1}{2} $.
\subsection{Velocity Saturation and Mobility Reduction}
Reducing the length of the transistor needs to include more features, such as mobility reduction, velocity saturation, and velocity overshoot for instance.
The mobility $ \mu_{eff} = v_{drift}/E_y $ , called the effective mobility, links the drift velocity to the longitudinal and vertical electric fields. AlGaN semiconductors exhibit saturation of the drift velocity
($ v_{drift} $) for electrons ($ v_{sat} $) when
the longitudinal electrical field $ E_y $ exceeds a
critical value $E_{crit}$, that also depends
on the vertical electrical field. In the effective mobility, normalizing the $ v_{drift} $ to $ v_{sat} $ and $ E_y $ to the $E_{crit}$ leads to an effective mobility $ u_{eff} $ \cite{Mangla}
\begin{equation}\label{mu1} 
u_{eff}=\frac{v_{drift}/v_{sat}}{E_y/E_{crit}}=\frac{\mu_{eff}}{\mu_x}=\frac{v}{e}
\end{equation}
where $e$ stand for the normalized electric field  ($e = E_y/E_{crit} $) and
$ v $ for the normalized velocity ($ v = v_{drift}/v_{sat} $). The mobility $ \mu_x$ is defined as $\mu_x= v_{sat}/E_{crit} $ and includes the effect of the vertical electrical field since the $ E_{crit} $ is also dependent on the vertical electrical field $ E_x $.
The normalized effective mobility to the low-field mobility $ \mu_0 $ is rewritten as 
\begin{equation}\label{mu}
u=\frac{\mu_{eff}}{\mu_0}=\frac{\mu_{eff}}{\mu_x} \frac{\mu_{x}}{\mu_0}=u_{eff} \times u_0,
\end{equation}
where $ u_0 $ is given by
\begin{equation}\label{eta}
u_0=\mu_{x}/\mu_0=[1+\theta(V_{GB}-V_T)]^{-1},
\end{equation}
Note that in (\ref{eta}), $ V_T $ is the charge threshold voltage which is not to be confused with the extrapolated threshold voltage, i.e. $V_{T0}$, and $ \theta $  corresponds to  the mobility reduction coefficient \cite{jazaeri2017free}. As already discussed in \cite{HEMT}, the drain current in HEMT is given by the drift-diffusion equation 
$ I_{DS}=\mu_{eff} W \left[Q_{ch}E_y+U_TdQ_{ch}/dy\right] $.
\begin{table}[t]
	\centering
	\renewcommand{\arraystretch}{1.5}
	\caption[List of symbols used in the model derivation.]{List of symbols used in the model derivations along the paper.}
	\label{Table2}
	\begin{tabular}{lll}
		\hline
		\textbf{Description} & \textbf{Symbol} & \textbf{Unit} \\
		\hline
		%		\textbf{Channel and Gate Charge Densities } & $ n_{ch} $, $ n_{G} $ & \SI {}{m^{-2}}  \\
		%		\textbf{Electron Charge } & $ q $ & C  \\
		\textbf{Thermal Voltage} & $ U_T $ & V  \\
		\textbf{Potential Profile and Surface Potential} & $ \psi $, $ \psi_s $ & V \\
		\textbf{Shift in the Quasi Fermi Level} & $V_{ch} $ & V \\
		%               \textbf{Bottom of the Conduction Band} & $ E_C $ & eV  \\
		%               \textbf{Top of the Valence Band } & $ E_V $ & eV  \\
		%               \textbf{Fermi Level} & $ E_F $ & eV  \\
		%               \textbf{First and Second Lowest Eigenstates} & $ E_0 $, $ E_1 $ & eV  \\
		%               \textbf{Input and Output Transconductances} & $ g_m $, $ g_{ds} $  & S  \\
		\textbf{Drain to Source Current} & $ I_{DS} $ & A  \\
		%               \textbf{Electric Field} & $ \mathscr{E} $ & V/m  \\  
		\textbf{Experimentally Determined Parameter} & $ u_1 $ & $ - $  \\
		%               \textbf{Electric Displacement Field} & $ D $ & C/m$ ^2 $  \\
		\textbf{Equivalent Barrier Capacitance} & $ C_b $ &  F/m$^2 $  \\
		\textbf{Body Factor} & $ \gamma $ &  V$^{1/2}$   \\
		\textbf{The Slope Factor} & $ n_q $ & $ - $  \\
		\textbf{Offset Voltage and Pinch-off Voltage} & $ V_A $, $ \psi_P $ & V  \\
		\textbf{Specific Charge } & $ Q_{SP} $ & C  \\
		\textbf{Specific Current} & $ I_{SP} $ &  A  \\
		\textbf{2D Density of States} & $ DoS_{2D} $ & $  m^{-2}V^{-1} $  \\
		\hline
	\end{tabular}
	\vspace{-0.4cm}
\end{table}
Normalizing the longitudinal electrical field, $ E(y) $, charge density, and current to $ E_{crit} $, $ Q_{sp} $, and $I_{sp} $ respectively,  the current can be written in the normalized from as 
\begin{equation}\label{ddn}
i=-u \left(\frac{2q_{ch}e}{\lambda_c}+\frac{dq_{ch}}{d\xi}\right),
\end{equation}
where $ \xi=y/L_G $, $ e=E_y/E_{crit} $ and $ \lambda_c $, called the velocity saturation parameter, is defined as $  \lambda_c =2U_T/E_cL_G $.
In a device operating in velocity saturated regime, the channel charge density at the drain approaches a saturated value $ q_{d,sat} $ which remains almost constant over the velocity saturated region, therefore $ dq_{ch}/d\xi =0 $. Relying on the current continuity, using piecewise linear velocity-field model (i.e. $ |e|u_{eff}=1 $ for $ |e|\geq 1 $ and $ u_{eff}=u_0 $ for $ |e|<1 $), and integrating (\ref{ddn}) over the velocity saturated part of the channel close to the drain, the normalized drain current in a velocity saturated device is given by $ i_{d,sat}=2u_0q_{d,sat}/\lambda_c $. On the other hand, relying on the current continuity along the channel, the drain current can be obtained from $ (\ref{current}) $ as\cite{Mangla}
\begin{equation}\label{dd}
i_{d,sat}=u_0(q^2_s+q_s-q^2_{d,sat}-q_{d,sat})=IC.
\end{equation}
For the sake of simplicity, we first propose to neglect the impact of the mobility reduction due to the vertical electric field, assuming $ u_0= 1 $. Later on, this non-ideal effect will be merged into the solution of the input characteristic. Introducing inversion coefficient $ IC $ to determine the channel inversion level of a HEMT as $ IC= I_{D,sat}/I_{sp} =i_{d,sat} $, the normalized source charge density in the saturation regime for a velocity saturated device is obtained form (\ref{dd}):  
\begin{equation}\label{nchsat1}
q_{s}=-\frac{qn_{ch}\mid_{s}}{Q_{sp}}=\frac{1}{2}\sqrt{\lambda^2_cIC^2+2\lambda_cIC+4IC+1}-\frac{1}{2}.\!\!
\end{equation}
It is worth mentioning that to some extent, \textit{naming 'inversion' and introducing IC to identify such a level of inversion in HEMT devices may seem unappropriated}. However, to make HEMT look like a generalized \textit{silicon MOSFET} we can still classify the operation regions of a HEMT as weak inversion (WI) for $ IC \leq 0.1 $, moderate inversion (MI) for $ 0.1 < IC \leq 10 $, and strong inversion (SI) for $ IC > 10 $, even though the channel is not resulting from an inversion process. 

Inserting relation (\ref{nchsat1}) into (\ref{x}) while imposing $ V_{ch}=0 $ leads to a  relationship between $ IC $ and $ V_{GB} $ in saturation mode. The relations (\ref{nchsat1}) and (\ref{x}) give an explicit solution of $ V_{GB} $ for a given IC. The general current–voltage relationship for a HEMT in saturation is independent of technological parameters and makes use of normalized variables. In the presence of mobility reduction due to the vertical field, IC can be replaced by $ u_0 $IC. Then using (\ref{x}), the large- and small-signal characteristics over a wide range of IC, from weak to strong 'inversion', can be fully captured by only six design parameters i.e. $ V_{T0} $, $ n_q $, $ I_{sp} $, $ \lambda_c $, $ V_T $, and $ \theta $.
On the other hand, differentiating (\ref{current}), the input transconductance in a HEMT is obtained 
\begin{equation}
g_m\!=\!\frac{\partial I_{DS}}{\partial V_{GB}}\!=\!I_{sp}\left[(2q_s+1)\frac{\partial q_s}{\partial V_{GB}}\!-\!(2q_d+1)\frac{\partial q_d}{\partial V_{GB}}\right]\!
\end{equation} 
where $ \partial q_s/\partial V_{GB} $ and $ \partial q_d/\partial V_{GB} $ are obtained from (\ref{x}) and are expressed by $\partial q_s/\partial V_{GB} \approx \partial q_d/\partial V_{GB} \approx 1/(2n_qU_T)  $. Therefore, an approximate value of $ g_m $ is given by 
\begin{equation}\label{gm}
g_m=I_{sp}(q_{s}-q_{d})/(n_qU_T).
\end{equation}
 Knowing $ g_m $, the best approach of interpreting the IC is through the transconductance efficiency, i.e. $ g_m/I_d $ versus IC  which is a key figure of merit widely used in CMOS circuit design: 
\begin{equation}\label{gm_I}
\begin{split}
\frac{g_mn_qU_T}{I_{DS}}&=\frac{q_{s}-q_{d}}{i}=\frac{1}{q_{s}+q_{d}+1}\\&=\frac{2}{\sqrt{(1+\lambda_cIC)^2+4IC}+1+\lambda_cIC}.
\end{split}
\end{equation}
Considering the relationship between the transconductance efficiency and IC allows for a proper evaluation of the design trade offs among gain, performance of analog design, transistor dimensions. This ratio reach the maximum in the weak 'inversion' ($ g_mn_qU_T/I_{DS} = 1$) where the drain current dependence is exponential versus $ V_{GB} $. However, moving toward to the strong inversion, the $ g_mn_qU_T/I_{DS} $ dependence versus IC becomes linear and therefore $ g_mn_qU_T/I_{DS} \approx 1/(\lambda_cIC)$ for large values of IC. 
%\section{$g_m$/$I$ based methodology for the design of CMOS analog circuits}
%The $g_m/I$ ratio plays a pivotal role in the design of amplifiers, filters, and various other analog circuits, influencing their performance metrics such as gain, bandwidth, linearity, and power consumption. In practical analog circuit design, achieving the desired $g_m/I$ ratio involves a strategic selection of transistor sizes and biasing currents tailored to the specific application. This ratio can also be determined experimentally through measurements on fabricated transistors, ensuring that the actual performance aligns with the intended design specifications. Equation (\ref{gm_I}) provides a fundamental link between the $g_m/I$ ratio and the intrinsic parameters of the transistor. Through analytical calculations using this equation, we derive a quantitative understanding of the transistor's behavior within amplification circuits.
Once the $g_m/I$ ratio is established, determining the optimal width-to-length ratio ($W/L$) of the transistor becomes straightforward. 
%For instance, to meet a predefined DC gain requirement $A_o$, given the Early voltage ($V_A$), the equation $A_o = -g_mV_A/I_{DS}$ facilitates the determination of the appropriate $W/L$ ratio necessary to achieve the desired gain. Similarly, in scenarios where the transition unity-gain frequency ($f_T$) needs to be calculated based on a load capacitance ($C_L$), the equation $f_T = (g_m/2\pi)/C_L$, combined with the known $g_m/I$ ratio, allows for the computation of the corresponding $W/L$ ratio.

%By leveraging the $g_m/I$ methodology, designers can effectively navigate the intricacies of HEMT-based analog circuit design, optimizing performance while meeting design objectives. This approach provides a systematic framework for translating design requirements into tangible transistor configurations, ensuring robust circuit functionality across various applications.
\section{Parameter Extraction Procedure}

This section describes a straightforward and precise method for parameter extraction for the simplified design-oriented EPFL HEMT model. This method necessitates six design parameters alongside some physical constants.

The initial step involves extracting $ n_q $ from the minimum of $ I_{DS}/g_mn_qU_T $ in weak inversion. It is worth noting that this ratio reaches the minimum value of 1 in weak inversion, leading to $ n_q=I_{DS}/g_mU_T$. This minimum occurs when the drain current dependence is exponential with respect to $ V_{GB} $.

Consequently, the slope factor can be extracted experimentally from the plot of $I_{DS}/g_mU_T $ versus $ I_{DS} $ in weak inversion. Similarly, the specific current $ I_{sp} $ can be determined from the same plot. Adjusting $ I_{sp} $ to match the asymptotic curvature of $ n_q=I_{DS}/g_mU_T$ to the experimental data allows to extract $ I_{sp} $.

Moving forward, $ \lambda_c $ is adjusted until the experimental data intersects the $1/\lambda_c IC$ line. Subsequently, the parameter $ n_q $ is modified to ensure that both the curves $g_m.n.U_T/I_{DS}$ and $ n_q $ intersect the asymptotic lines simultaneously.

Then, $V_T$ and $V_{T0}$ are utilized to fit the $I_{DS}-V_{GB}$ curve near the threshold and the rising part of $g_m$. Finally, $ \theta $ and other variables are fine-tuned to fit the $I_{DS}-V_{GB}$ curve at higher $V_{GB}$ and the falling part of the $g_m$ curve.
\begin{table}[b]
\vspace{-0.5cm}
	\centering
	\renewcommand{\arraystretch}{1.5}
	\caption[List of extracted parameters for both long and short channel GaN HEMTs.]{List of extracted parameters for both long and short channel GaN HEMTs.}
	\label{Table2}
	\begin{tabular}{llll}
		\hline
		Parameter & Unit &\textbf{$L = 3.0 \mu m$} & \textbf{$L = 200 nm$} \\
		\hline
		\textbf{$ I_{sp} $} & \textbf{$A$} & $ 2.6 \mu $ & $38.8 \mu $  \\
		\textbf{$n$ } & \textbf{$-$} & $ 1.6 $ $  $ & $2.75$ \\
		\textbf{$ L_{sat} $ } & \textbf{$m$} & $335 n $ & $135 n $ \\
  		\textbf{$ \lambda_{c} $ } & \textbf{$-$} & $111.7 m $ & $675 m $ \\
      		\textbf{$ V_{T0} $ } & \textbf{$V$} & $-2.8 $ & $-3.0 $ \\
              		\textbf{$ \theta $ } & \textbf{$-$} & $0.26 $ & $0.11 $ \\
                      		\textbf{$ V_T $ } & \textbf{$V$} & $-2.6 $ & $-1.10 $ \\
		\hline
	\end{tabular}
	\vspace{-0.4cm}
\end{table}
\begin{figure*}[t!]
  \includegraphics[width=\linewidth]{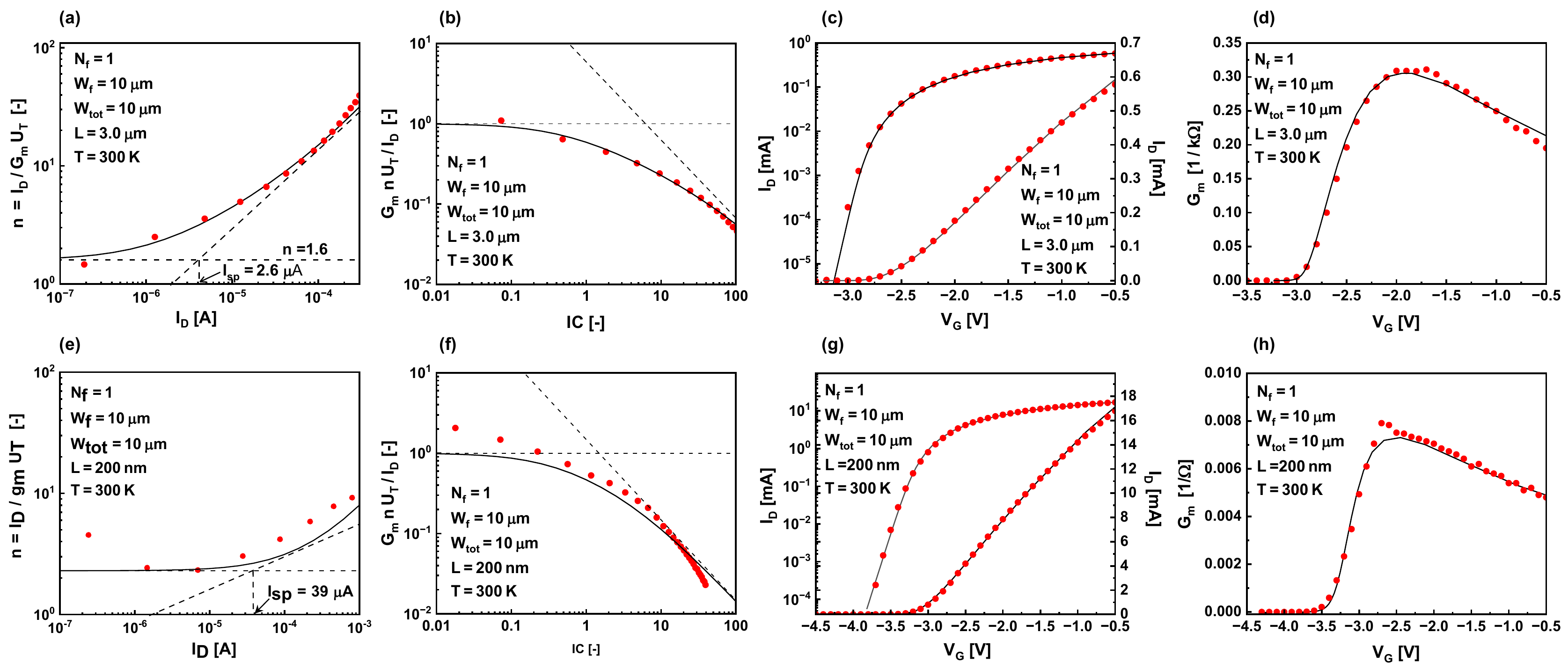}
  \caption{The comparison of measured data and model for short and long GaN devices. The symbols represent the measurements, while the solid line corresponds to the simplified EPFL model. $N_f$ corresponds to the number of fingers, $W_f$ is the width of each finger, and $W_{tot}$ represents the total width calculated as $N_f \times W_f$. Here, $L$ denotes the gate length, while $T$ corresponds to temperature. 
  The drain-to-source voltage is maintained at 1V, ensuring devices operate in the linear mode of operation. (a, e) illustrate the $n$ factor versus drain current. (b, f) depict the $G_m/I$ versus IC for short and long channels. (c, g) present the input characteristics in both linear and logarithmic scales. (d, h) display the $G_m$ versus $V_G$ for both channel lengths.}
  \label{fig:Fig1}
  \vspace{-0.4cm}
\end{figure*}
%\section{Device Fabrication and Experimental results}
%In our experiments, the AlInN/GaN transistor structures were grown on sapphire substrates using MEMOCVD® technique developed by Sensor Electronic Technology Inc….
%Room temperature measurements were performed on devices fabricated and are reported in the next section.
%The electrical input and output characteristics of the devices were measured in a dark and screening vacuum chamber at RT on a long pMOS device with W/L=.., in linear and logarithmic scales
%Figure 3 shows transfer characteristics varying with operation temperature.
%Figure …. shows the output characteristics of the devices with the gate geometry of 10-$ \mu $ m2$ \times $2 and gate length of 10-$ \mu $m which are evaluated at room temperature, and the transition between both temperatures.

%DC characteristics are obtained from on-wafer measurements using an ..... probe station at room temperature.  The results were acquired with a Keysight B1500A semiconductor device
%parameter analyzer. Using this set-up, we measured transfer characteristics in the linear (VDB = 10 mV) and the saturation regions (VDB = 0.9 V), as well as output characteristics for different gate voltages.

\section{Results and Discussion}
The model has been validated experimentally on GaN-on-Si HEMTs fabricated in IMEC following \cite{8993582}. The epitaxial stack grown on 200 mm HR Si wafers by MOCVD used in this study is composed of: 5nm in-situ SiN passivation layer, 15nm AlGaN barrier, 1nm AlN spacer, 300nm GaN channel, on top of 1um C-doped back barrier and 1um transition layers.

In Figure 2, the slope factors ($n_q$) are derived from the plateau of $I_D/(G_m·U_T)$ versus $I_D$ in weak inversion (WI) for both long-channel ($L_G = 3.0 \mu m$), shown in Fig. 2.a, and short-channel ($L_G = 200 nm$), depicted in Fig. 2.e. 
This essential extraction process captures the behavior of the drain current and transconductance, offering profound insights into device performance across different channel lengths.

Additionally, the normalized transconductance efficiency, $ G_mnU_T /I_D $, is plotted versus the inversion coefficient $ IC $. 
Figures 2.b and 2.f serve to corroborate the robustness of the $G_m/I_D$ design methodology. In particular, from the bare intersection of the asymptotes in weak and strong 'inversion', see dashed lines in Figure 2, the key specific current parameter $I_{sp}$ for both short and long channel devices is obtained.

Figures 2c and 2g present the $I_D$ versus $V_G$ characteristics for both long and short channel GaN HEMTs, depicted in both linear and logarithmic scales. These plots demonstrate that the simplified EPFL model accurately captures the input transfer behaviour across different channel lengths. Whether in the linear or logarithmic scale, the model accurately reflects the device characteristics, underscoring its efficacy in modelling both long and short channel devices. 

In summary, Figures 2d and 2h show a detailed comparison of the transconductance ($G_m$) against $V_G$ for both long and short channel GaN HEMTs. The remarkable fidelity with which the simplified EPFL model captures both $G_m$ and its peak values across varying gate voltages is confirmed. This robust agreement between model predictions and experimental data highlights the model's capability to accurately represent the transconductance behavior of GaN HEMTs. The extracted parameters for both long and short channel GaN HEMTs are listed in Table II. 
This validation underscores the simplified model effectiveness to model HEMTs using only a few parameters and enables a design and optimization of GaN-based devices in a more traditionnal way.
\section{Conclusion}
This paper introduces a simplified and design-oriented adaptation of the EPFL HEMT model, with a specific focus on the normalized transconductance-to-current characteristic  and IC. The research delves into GaN HEMT technology and modeling, aiming to provide a precise model for the electrical behavior of these devices using only a few parameters.
Validation is conducted by comparing measured transfer characteristics of GaN HEMTs at room temperature across a broad range of IC values. This study provides valuable guide in the effective design of circuits employing HEMTs.

% conference papers do not normally have an appendix

% use section* for acknowledgment

% trigger a \newpage just before the given reference
% number - used to balance the columns on the last page
% adjust value as needed - may need to be readjusted if
% the document is modified later
%\IEEEtriggeratref{8}
% The "triggered" command can be changed if desired:
%\IEEEtriggercmd{\enlargethispage{-5in}}

% references section

% can use a bibliography generated by BibTeX as a .bbl file
% BibTeX documentation can be easily obtained at:
% http://mirror.ctan.org/biblio/bibtex/contrib/doc/
% The IEEEtran BibTeX style support page is at:
% http://www.michaelshell.org/tex/ieeetran/bibtex/
%\bibliographystyle{IEEEtran}
% argument is your BibTeX string definitions and bibliography database(s)
%\bibliography{IEEEabrv,../bib/paper}
%
% <OR> manually copy in the resultant .bbl file
% set second argument of \begin to the number of references
% (used to reserve space for the reference number labels box)

%\bibliographystyle{IEEEtran}
%\bibliography{IEEEabrv,biblio}

\begin{thebibliography}{1}
\bibitem{HEMT}
F.~Jazaeri and J.-M. Sallese, ``{Charge-based EPFL HEMT Model},'' \emph{IEEE
  Transactions on Electron Devices}, vol.~66, no.~3, pp. 1218--1229, 2019.

\bibitem{EKV1}
{Christian C. Enz, Eric A. Vittoz}, \emph{{Charge-Based MOS Transistor
  Modeling: The EKV Model for Low-Power and RF IC Design}}.\hskip 1em plus
  0.5em minus 0.4em\relax Wiley, 2006.

\bibitem{tran}
F.~Jazaeri, M.~Shalchian, and J.-M. Sallese, ``{Transcapacitances in EPFL HEMT
  Model},'' \emph{IEEE Transactions on Electron Devices}, vol.~67, no.~2, pp.
  758--762, 2020.

\bibitem{9137643}
M.~Allaei, M.~Shalchian, and F.~Jazaeri, ``{Modeling of Short-Channel Effects
  in GaN HEMTs},'' \emph{IEEE Transactions on Electron Devices}, vol.~67,
  no.~8, pp. 3088--3094, 2020.

\bibitem{jazaeri2017free}
F.~Jazaeri, A.~Pezzotta, and C.~Enz, ``{Free Carrier Mobility Extraction in
  FETs},'' \emph{IEEE Transactions on Electron Devices}, vol.~64, no.~12, pp.
  5279--5283, 2017.

\bibitem{EG}
{Michael E. Levinshtein (Editor), Sergey L. Rumyantsev (Editor), Michael S.
  Shur (Editor)}, \emph{{Properties of Advanced Semiconductor Materials: GaN,
  AIN, InN, BN, SiC, SiGe}}.\hskip 1em plus 0.5em minus 0.4em\relax John Wiley and Sons, Inc., New York, 2001.

\bibitem{linearization}
J.-M. Sallese, M.~Bucher, F.~Krummenacher, and P.~Fazan, ``{Inversion Charge
  Linearization in MOSFET Modeling and Rigorous Derivation of the EKV Compact
  Model},'' \emph{Solid-State Electronics}, vol.~47, no.~4, pp. 677--683, 2003.

\bibitem{Mangla}
A.~Mangla, ``{Modeling Nanoscale Quasi-Ballistic MOS Transistors: A Circuit
  Design Perspective},'' 2014.

\bibitem{8993582}
U.~Peralagu, A.~Alian, and e.~a. Putcha, ``{CMOS-compatible GaN-based Devices
  on 200mm-Si for RF Applications: Integration and Performance},'' pp.
  17.2.1--17.2.4, 2019.

\end{thebibliography}

\end{document}